\documentclass[showpacs,preprintnumbers,amsmath,amssymb]{revtex4}
\usepackage{amssymb}
\usepackage{graphicx}% Include figure files
\usepackage{dcolumn}% Align table columns on decimal point
\usepackage{bm}% bold math
\usepackage{textcomp}
\usepackage{overpic}

\newcommand\oprod[2]{\ensuremath{|#1\rangle\langle#2|}}

\begin{document}
\title{Decoy-state quantum key distribution with biased-bases revisited}
\author{Zong-Wen Yu$ ^{1,2}$, Yi-Heng Zhou$ ^{1,3}$,
and Xiang-Bin Wang$ ^{1,3,4\footnote{Email
Address: xbwang@mail.tsinghua.edu.cn}\footnote{Also at Tsinghua University, Center for Atomic and Molecular Nanosciences}}$}

\affiliation{ \centerline{$^{1}$State Key Laboratory of Low
Dimensional Quantum Physics, Tsinghua University, Beijing 100084,
People's Republic of China}\centerline{$^{2}$Data Communication Science and Technology Research Institute, Beijing 100191, People's Republic of China}\centerline{$^{3}$ Synergetic Innovation Center of Quantum Information and Quantum Physics, University of Science and Technology of China}\centerline{  Hefei, Anhui 230026, People's Republic of China
 }\centerline{$^{4}$ Shandong
Academy of Information and Communication Technology, Jinan 250101,
People's Republic of China}}

%%%%%%%%%%%%%%%%%%%%%%%%%%%%%%%%%%%%%%%%%%%%%%%%%%%%%%%%%%%%%%%%%%%
%%%%%%%%%%%%%%%%%%%%%%%%%%%%%%%%%%%%%%%%%%%%%%%%%%%%%%%%%%%%%%%%%%%
%%%%%%%%%%%%%%%%%%%%%%%%% Abstract %%%%%%%%%%%%%%%%%%%%%%%%%%%%%%%%
\begin{abstract}
In order to improve the key rate of the decoy-state method, we need
to jointly study yields of different bases. Given the delicate fact
that pulses of the same preparation state can have different
counting rates if they are measured in different bases, for example,
those vacuum pulses and those single-photon pulses, existing results
of decoy-state quantum key distribution using biased bases are
actually flawed by assuming that they are equal. We fix this flaw
through using the idea that yields of pulses prepared in different
bases are same provided that they are prepared in the same state and
also they are measured in the same basis, for example, those
single-photon pulses prepared in different bases but measured in the
same basis.  Based on this, we present correct formulas for the
decoy-state method using biased bases. Taking the effects of
statistical fluctuations into account, we then numerically study the
key rates of different protocols with all parameters being fully
optimized. Our result confirms the prior art conclusion that
decoy-state method using biased bases can have advantage to the
symmetric protocol with unbiased bases. We obtain high key rates of
our 4-intensity protocol (two in X bases and two in Z bases) without using any vacuum source.

%Furthermore, the methods with the vacuum source or not are nearly equal to each other.
\end{abstract}

%%%%%%%%%%%%%%%%%%%%%%%%%%%%%%%%%%%%%%%%%%%%%%%%%%%%%%%%%%%%%%%%%%%
%%%%%%%%%%%%%%%%%%%%%%%%%%%%%%%%%%%%%%%%%%%%%%%%%%%%%%%%%%%%%%%%%%%
%%%%%%%%%%%%%%%%%%%%%%%%%%%%%%%%%%%%%%%%%%%%%%%%%%%%%%%%%%%%%%%%%%%

\maketitle

%%%%%%%%%%%%%%%%%%%%%%%%%%%%%%%%%%%%%%%%%%%%%%%%%%%%%%%%%%%%%%%%%%%
%%%%%%%%%%%%%%%%%%%%%%%%%%%%%%%%%%%%%%%%%%%%%%%%%%%%%%%%%%%%%%%%%%%
%%%%%%%%%%%%%%%%%%%%%%%%%%%%%%%%%%%%%%%%%%%%%%%%%%%%%%%%%%%%%%%%%%%
%%%%%%%% Introducation & Motivation %%%%%%%%%%%%%%%%%%%%%%%%%%%%%%%

\section{Introduction}\label{SecIntro}
Quantum key distribution (QKD) is one of the most successful applications of quantum information processing. It can help two remote parties, commonly noted as Alice and Bob, to set up the unconditionally secure key. The security of QKD is based on the fundamental laws of quantum physics~\cite{BB84,GRTZ02}, rather than unproven of computational complexity assumptions.

In practice, imperfect single-photon sources are used in most of the real setups of QKD~\cite{BB84,GRTZ02,ILM}. Such implementations, in principle, suffer from the photon-number-splitting (PNS) attack~\cite{PNS,PNS1}. The decoy-state method~\cite{rep,H03,wang05,wang06,LMC05,LMC051,AYKI,haya,peng,wangyang,njp} and some other methods~\cite{UrsinNP2007,scran,kko} can be used for unconditionally secure QKD even if Alice only uses the imperfect sources~\cite{PNS,PNS1}.

The central issue in the decoy-state method is to faithfully estimate the yield of single-photon pulses. If we do this estimation in each bases separately, i.e., only use those observed outcome of pulses prepared and measured in the same basis, we can obtain two different values of yields, one for the $X$ basis, one for the $Z$ basis. In such a way, results will be faithful but the key rate are not optimized. We can improve the key rate by using the observed results in different bases {\em jointly}, however, there are delicate points we must pay attention to.

In fact, pulses of the same preparation state can have different counting rates if they are measured in different bases, for example, those vacuum pulses and those single-photon pulses. This point is not considered in all previous biased-basis methods~\cite{LoJC2005,MaSRep2013,gm}. In this paper, we point out that yields of pulses prepared in different bases are same if they are prepared in the same state and measured in the same basis, for example, those single-photon pulses. This is to say, we can treat those measurement outcome from different preparation bases {\em jointly} provided that they are measured in the same basis. Based on this, we present correct formulas for the decoy-state method using biased bases. We also consider the effects of statistical fluctuations.

In order to make a good estimation of the final key rate with the decoy-state methods, we need to find out the lower and upper bounds of the yield $s_0$ caused by the vacuum state first. These bounds can be easily obtained if we assume that Alice can prepare a vacuum source. However, in practice, the different intensities are usually generated with an intensity modulator, which has a finite extinction ratio. So it is usually difficult to create a perfect vacuum state in decoy-state QKD experiments~\cite{peng}. In this paper, we also make a study on the decoy-state method without using vacuum. This is particularly important for the practical implementations.

Regarding the yield $s_0$ of the vacuum state as a common variable, the final key rate is a function of $s_0$. There is a possible region for this values. To extract the final secret key, we should find out the worst-case key rate over the whole region of $s_0$ values. While in some previous works~\cite{MaSRep2013,gm}, they straightly use the lower bound of $s_0$ to calculate the final key rate. Actually, in some cases shown in Sec.~\ref{sec:Simulation}, we find that the smallest value of the final key rate does not appear with the lower bound of $s_0$. In this case, if we simple mindedly calculate the final key rate with the lower bound of $s_0$, the security of the decoy-state method will not be guaranteed.

The rest of this paper is organized as follows. In Sec.~\ref{sec:Protocol}, we review some delicate points of the decoy-state method. Subsequently, we show the correct formulas to estimate the lower bound of $s_1$ and the upper bound of $e_1$ with the effect of statistical fluctuations being taken into account. After that, in Sec.~\ref{sec:Simulation}, we numerically study the optimized key rates of different protocols.
The article is ended in Sec.~\ref{sec:Conclusion} with a concluding remark.

\section{Some delicate points of the decoy-state method and the correct formulas}\label{sec:Protocol}
Consider the case we use two bases for state preparation and measurement, the {\em Z} basis and the {\em X} basis.
We denote $s_{1,\alpha}^\omega$ for the yield of those single-photon pulses which are prepared in the $\alpha$ basis and measured in the $\omega$ basis. Note that the state of single-photon pulses prepared in the $Z$ basis is same with state of single-photon pulses prepared in the $X$ basis. The density matrix of both states are simply $I/2$.  We also denote $s_0^\omega$ as the yield of those vacuum pulses which are measured in the $\omega$ basis. The delicate point is that even in the asymptotic case
\begin{equation}
  s_0^Z\not=s_0^X, \label{d1}
\end{equation}
and
\begin{equation}
  s_{1,\alpha}^Z\not=s_{1,\beta}^X. \label{d2}
\end{equation}
These are different from the assumptions in prior art works such as~\cite{MaSRep2013} where one simply chose to do the decoy-state analysis {\em only} in $Z$ basis and then just used the value of  $s_{1,Z}^Z$ for the value $s_{1,X}^X$.
The reason $s_{1,Z}^Z\not=s_{1,X}^X$ is  simply due to the mismatch of detection efficiencies and dark counts in different bases.
Such mismatch can come from either imperfect control two the devices inside Labs, or Eve's attack~\cite{DEM1,DEM2,FSA,TimeShift}.
%If we disregard the above inequalities~(\ref{d1}, \ref{d2}), the final key will be insecure under many types of attacks, such as the faked-state
%attack~\cite{FSA} and the time-shift attack~\cite{DEM1,TimeShift}.
If we do the decoy-state study in each basis separately with unbiased bases, we can obtain the lower bound values in each bases separately for $s_{1,Z}^Z$ and $s_{1,X}^X$. The results are faithful but the key rate will be low. Note that here we must use $s_0^Z$ and $s_0^X$ separately in calculating $s_{1,Z}^Z$ and $s_{1,X}^X$.

Here we have a better treatment and we can still study the decoy-state method {\em jointly} in different bases.
For this goal, {\em we shall use the observed number of counts of pulses prepared in one basis but measured in another basis.}
In particular, we have the following elementary equalities
\begin{equation}
  s_{1,Z}^Z= s_{1,X}^Z, \label{c1}
\end{equation}
and
\begin{equation}
  s_{1,Z}^X= s_{1,X}^X. \label{c2}
\end{equation}
Here, $s_{1,\alpha}^\omega (\alpha,\omega =Z,X)$ is the yield of those single-photon pulses which are prepared in the $\alpha$ basis and measured in the $\omega$ basis. Eqs.~(\ref{c1}, \ref{c2}) are the main idea of this paper. Given these equations, we {\em don't} have to study the decoy-state method completely separately in each bases. For example, consider a protocol using a vacuum source $O$, one source $X_1$ in the $X$ basis and two sources $\{Z_{j}|j=1,2\}$ in the $Z$ basis. Observing $S_O^\omega$ and $\{S_{Z_j}^{\omega}|j=1,2\}$ we can formulate the yield of single-photon pulses measured in the $Z$ basis and also the yield of single-photon pulses measured in the $X$ basis. Here $S_O^\omega$ is the yield of the vacuum source measured in the $\omega$ basis and $S_{Z_j}^{\omega}$ is the yield of source $Z_j$ measured in the $\omega$ basis. Using the observed values $\{S_{Z_j}^{\omega}|j=1,2\}$ and $S_O^\omega$. Asymptotically we have
\begin{equation}\label{c5}
  s_{1,Z}^{\omega}\geq s_{1,Z}^{\omega,L} = \frac{a_{2,Z_2} S_{Z_1}^{\omega}-a_{2,Z_1}S_{Z_2}^{\omega} - A_{Z_1,Z_2}^{0,2} S_{O}^{\omega}} {A_{Z_1,Z_2}^{1,2}},
\end{equation}
where $A_{Z_1,Z_2}^{0,2}=a_{0,Z_1} a_{2,Z_2}-a_{0,Z_2}a_{2,Z_1}$, $A_{Z_1,Z_2}^{1,2}=a_{1,Z_1} a_{2,Z_2}-a_{1,Z_2}a_{2,Z_1}$ and $a_{k,Z_j}$ are non-negative parameters of sources $Z_j$ with $\rho_{Z_k}=\sum_{k} a_{k,Z_j}\oprod{k}{k}$. Here $\omega$ can take both $X$ and $Z$.  Given Eqs.~(\ref{c1}, \ref{c2}), the error rate of single-photon pulses prepared and measured in the $X$ basis is
\begin{equation}\label{e5}
  e_1^X \leq e_{1}^{X,U}=\frac{T_{X_1}^{X}-a_{0,X_1} s_0^X/2}{a_{1,X_1} s_{1,X}^{X,L}}= \frac{T_{X_1}^{X}-a_{0,X_1}S_O^X/2}{a_{1,X_1}s_{1,Z}^{X,L}},
\end{equation}
where $T_{X_1}^{X}$ are error yield of the source $X_1$ measured in the $X$ basis, $a_{k,X_1}$ are non-negative parameters of source $X_1$ with $\rho_{X_1}=\sum_{k} a_{k,X_1}\oprod{k}{k}$ and the lower bound $s_{1,Z}^{X,L}$ is given by Eq.~(\ref{c5}) already. Note that our formulas Eqs.~(\ref{c5}, \ref{e5}) are unconditionally correct under whatever situation, e.g.,  detection efficiency mismatch, no matter it comes from the imperfect control in side Lab. or Eve.'s attack outside Lab~\cite{DEM1,DEM2,FSA,TimeShift}.

Most generally, Alice can use several different sources. We assume
Alice prepares the vacuum source $O$, $2$ non-vacuum sources in the
$Z$ basis, $Z_1,Z_2$ and $2$ different non-vacuum sources in the $X$
basis $X_1,X_2$ with probabilities $p_O$, $p_{Z_1}$, $p_{Z_2}$ and
$p_{X_1}$, $p_{X_2}$,respectively.
%Here, $Z_l (l=1,2)$ and $X_r (r=1,2)$
%are prepared in the $Z$ and $X$ bases respectively.
The density
matrices of these non-vacuum sources in photon number space are
denoted as follows
\begin{equation}\label{eq:rho}
  \rho_{\alpha_{l}}=\sum_{k} a_{k,\alpha_j}\oprod{k}{k},
\end{equation}
where $a_{k,\alpha_l}$ are non-negative parameters and $\alpha$ can
be the $Z$ and $X$ bases. Meanwhile, we introduce the following very
important conditions for two different sources ${\alpha_{1}}$ and
${\alpha_{2}}$
\begin{equation}\label{eq:cond}
  \frac{a_{k,\alpha_{2}}}{a_{k,\alpha_{1}}}\geq \frac{a_{1,\alpha_{2}}}{a_{1,\alpha_{1}}}\geq \frac{a_{0,\alpha_{2}}}{a_{0,\alpha_{1}}},
\end{equation}
for all $k\geq 2$. We denote ${\alpha_{1}}\prec {\alpha_{2}}$ when
the sources ${\alpha_{1}}$ and ${\alpha_{2}}$ fulfill the relations
presented in Eq.(\ref{eq:cond}). Imperfect sources used in practice
such as the coherent state source, the heralded source out of the
parametric-down conversion, satisfy the above conditions.

In the protocol, we also assume that Bob measures the received pulses in the $Z$ and $X$ bases with probabilities
$q^{Z}$ and $q^{X}$ respectively. After the preparation and measurement of $N_t$ pulses, Alice and Bob obtain the observable
$N_{\alpha_j}^{\omega}$ and $M_{\alpha_j}^{\omega}$ which are the number of successful counts and error counts when Alice sends the
pulses from source ${\alpha_j}$ and Bob measures them in the $\omega$ basis (when preparation basis and measured basis are different,
we do not need error counts). Here $\alpha$ and $\omega$ can take both $X$ and $Z$.
We also denote $S_{\alpha_j}^{\omega}$ and $T_{\alpha_j}^{\omega}$ as the yield and error yield respectively with
$S_{\alpha_j}^{\omega}={N_{\alpha_j}^{\omega}}/(p_{\alpha_j}q^{\omega} N_t)$
and $T_{\alpha_j}^{\omega}={M_{\alpha_j}^{\omega}}/({p_{\alpha_j} q^{\omega} N_t})$. As shown, we have the normalized
relations $p_{O}+\sum_{l=1}^{2}p_{Z_l}+\sum_{r=1}^{2}p_{X_r}=1$ and $q^X+q^Z=1$.

Following the GLLP security analysis~\cite{ILM}, the final key rate for the source $\rho_{Z_l}$ is given by
\begin{equation}\label{eq:RFormula}
  R= p_{Z_l} q^Z\left\{a_{1,Z_l}s_1^Z [1-H(e_1^{p,Z})]-f_e S_{Z_l}^{Z} H(E_{Z_l}^{Z})\right\},
\end{equation}
if the source is not used for error test. Here $p_{Z_l} q^Z$ is the raw data sift factor, including the basis-sift factor $q^Z$ and the signal-state ratio $p_{Z_l}$; $s_1^Z$ and $e_1^{p,Z}$ are the yield and phase error rate of the single-photon state measured in the $Z$ basis; $f_e$ is the efficiency factor of the error-correction method used; $S_{Z_l}$ and $E_{Z_l}$ are the yield and quantum bit error rate of the source $Z_l$ measured in the $Z$ basis; $H(x)=-x \log_2(x)-(1-x)\log_2(1-x)$ is the binary Shannon entropy function. The phase error rate $e_1^{p,Z}$ can be estimated from the error rate in the $X$ basis while it can not be measured directly.

In any real experiment, the total pulses sent by Alice is finite. In order to extract the secret final key, we have to consider the effect of statistical fluctuations caused by the finite-size. In this case, yields of the same state out of different sources are not always rigorously equal to each other, i.e. $s_{k,\alpha_{1}}^{\omega}\neq s_{k,\alpha_{2}}^{\omega}$. Here $s_{k,\alpha_j}^{\omega}$ is the yield of $k$-photon pulses prepared from source $\alpha_j$ and measured in the $\omega$ basis. To obtain the lower bound of $s_{1}^{\omega}$, one can implement the idea of Ref.~\cite{njp}, i.e. using the averaged yield of a specific state from different sources. As shown, one can introduce the averaged value for the yield of $k$-photon pulses from all sources that are {\em prepared in the same state and measured in the same basis}.
 We define
 %\begin{equation}
 %\langle{s}_{k,\alpha}^{\omega}\rangle
 %= \frac{1}{c_{k,\alpha}^{\omega}}\sum_{{j}} p_{\alpha_j} a_{k,\alpha_j} s_{k,\alpha_j}^{\omega}, \quad (k\geq 1), \label{eq:Msk}
 %\end{equation}
\begin{eqnarray}
  \langle{s}_{k,\alpha}^{\omega}\rangle &=& \frac{1}{c_{k}^{\omega}}\sum_{{l}} p_{Z_l} a_{k,Z_l} s_{k,Z_l}^{\omega}  \nonumber \\
  & & + \frac{1}{c_{k}^{\omega}}\sum_{{r}} p_{X_r} a_{k,X_r} s_{k,X_r}^{\omega},  \quad (k=0,1); \label{eq:Ms01} \\
  \langle{s}_{k,\alpha}^{\omega}\rangle &=& \frac{1}{c_{k,\alpha}^{\omega}}\sum_{{j}} p_{\alpha_j} a_{k,\alpha_j} s_{k,\alpha_j}^{\omega}, \quad (k\geq 2), \label{eq:Msk}
\end{eqnarray}
where $c_{k}^{\omega}= \sum_{{l}} p_{Z_l} a_{k,Z_l} + \sum_{{r}}
p_{X_r} a_{k,X_r}(k=0,1)$ and $c_{k,\alpha}^{\omega} =\sum_{{j}}
p_{\alpha_j} a_{k,\alpha_j} (k\geq 2)$. With Eq.~(\ref{eq:Ms01}), we
have $\langle s_{k,X}^{\omega}\rangle =\langle
s_{k,Z}^{\omega}\rangle$ for $k=0,1$. Therefore we shall omit
subscript $\alpha$ there for $k=0,1$ and use notations $\langle
s_{0}^{\omega}\rangle,\;\langle s_{1}^{\omega}\rangle$ for $\langle
s_{0,\alpha}^{\omega}\rangle,\; \langle
s_{1,\alpha}^{\omega}\rangle$ for simplicity. Also, we {\em define}
quantity
\begin{equation}\label{cons1}
  \langle{S}_{\alpha_l}^{\omega}\rangle= \sum_{k=0}^\infty a_{k,\alpha_l}\langle{s}_{k,\alpha}^{\omega}\rangle.
\end{equation}

Considering the relations in Eq.~(\ref{cons1})  prepared in the
$\alpha$ basis and measured in the $\omega$ basis, we can lower
bound $\langle s_{1,\alpha}^{\omega}\rangle$ for a given value of
$\langle s_0^{\omega}\rangle$ with the following
equations~\cite{wangyang}
\begin{equation}\label{ls1}
\langle s_{1,\alpha}^{\omega}\rangle \geq \langle
s^{\omega,L}_1\rangle =\max_{\alpha=Z,X}\left[\langle
s_{1,\alpha}^{\omega,L}\rangle(\langle s_0^\omega\rangle)\right]
\end{equation}
and
\begin{equation}\label{eq:s1sta}
  \langle s_{1,\alpha}^{\omega,L}\rangle (\langle s_0^\omega\rangle)
  = \frac{a_{2,\alpha_{2}} \underline{S}_{\alpha_{1}}^{\omega} -a_{2,\alpha_{1}} \overline{S}_{\alpha_{2}}^{\omega}
  -A_{\alpha_{1}\alpha_{2}}^{0,2} \langle s_{0}^{\omega}\rangle }{A_{\alpha_{1}\alpha_{2}}^{1,2}},
\end{equation}
where $A_{\alpha_{1}\alpha_{2}}^{0,2}=a_{0,\alpha_{1}}
a_{2,\alpha_{2}} -a_{0,\alpha_{2}} a_{2,\alpha_{1}}$,
$A_{\alpha_{1}\alpha_{2}}^{1,2}=a_{1,\alpha_{1}} a_{2,\alpha_{2}}
-a_{1,\alpha_{2}} a_{2,\alpha_{1}}$ and
\begin{equation}\label{eq:SlLU}
  \underline{S}_{\alpha_j}^{\omega}=S_{\alpha_j}^{\omega}/(1+{\delta}_{\alpha_j}^{\omega}), \quad
  \overline{S}_{\alpha_j}^{\omega}=S_{\alpha_j}^{\omega}/(1-{\delta}_{\alpha_j}^{\omega}).
\end{equation}
By using the multiplicative form of Chernoff
bound~\cite{CurtyNC2014,PanPRL2014}, with a fixed failure
probability $\epsilon$, we can give an interval of
$\langle{S}_{\alpha_j}^{\omega}\rangle$ with the observable
${S}_{\alpha_j}^{\omega}$, $ [\underline{S}_{\alpha_j}^{\omega},
\overline{S}_{\alpha_j}^{\omega}]$, which can bound the value of
$\langle{S}_{\alpha_j}^{\omega}\rangle$ with a probability at least
$1-\epsilon$. Similarly, we can also define $\overline{T}_{\omega_j}^{\omega}$ with the observable $T_{\omega_j}^{\omega}$. Note that $\langle s_1^{\omega,L}\rangle$ is actually a
function of $\langle s_0^{\omega}\rangle$.

With the mean values $\langle s_{1}^{\omega,L}\rangle$ defined in Eq.~(\ref{ls1}), the lower bounds of $s_{1,Z_l}^{Z}$ and $s_{1,X_r}^{X}$ can be estimated by
\begin{equation}\label{eq:s1L}
  s_{1,Z_l}^{Z,L}= \langle s_{1}^{Z,L} \rangle (1-\delta_{Z_l}), \quad s_{1,X_r}^{X,L}= \langle s_{1}^{X,L} \rangle (1-\delta_{X_r}),
\end{equation}
where $\delta_{Z_l}=\lambda/\sqrt{N_{1,Z_l}^{Z} \langle s_{1,Z}^{Z,L}\rangle}$ with $\lambda=\sqrt{-2\ln \epsilon}$ and $\delta_{X_r}=\lambda/\sqrt{N_{1,X_r}^{X} \langle s_{1,X}^{X,L}\rangle}$. Here and after, we define
\begin{equation}\label{eq:N1}
  N_{k,\alpha_j}^{\omega}=a_{k,\alpha_j}p_{\alpha_j}q^{\omega}N_t,
\end{equation}
as the number of $k$-photon pulses prepared in source $\alpha_j$ and measured in the basis $\omega$.

In order to estimate the final key rate, we also need the upper bound of the error rate $e_{1}^{X}$. Similarly, we have
\begin{equation}\label{eq:e1sta}
  e_{1,X_1}^{X} \leq e_{1,X_1}^{X,U}=\frac{{T}_{X_1}^{X}-a_{0,X_1}
  \langle s_0^X\rangle (1-\delta_{0,X_1}^{X}) /2}{a_{1,X_1} s_{1,X_1}^{X,L}},
\end{equation}
where $\delta_{0,X_1}^{X}=\lambda/\sqrt{N_{0,X_1}^{X} \langle
s_{0}^{X} \rangle}$.

If the key size is infinite, the phase-flip error rate for
single-photon counts in $Z$ basis is simply $e_{1}^{p,Z}=e_{1}^{X}$.
In a finite-key-size case, we can apply the large data size
approximation of the random sampling method~\cite{FungPRA2010} to
upper bound the phase error rate $e_{1}^{p,Z}$ of single-photon
pulses prepared and measured in the $Z$ basis with the failure
probability $\epsilon$
\begin{equation}\label{eq:e1pZ}
  e_{1}^{p,Z}\leq e_1^{p,Z,U}= e_{1,X_1}^{X,U} + \theta_{Z}^{X},
\end{equation}
where $\theta_{Z}^{X}=\sqrt{n_{\theta}/d_{\theta}}$ with
$d_{\theta}=\frac{(1-g_{X})g_{X}\ln 2}{2(1-e_1)e_1}$,
$n_{\theta}=-\log[\epsilon \sqrt{e_1(1-e_1)n_{X}
n_{Z}/(n_{X}+n_{Z})}]/(n_{X}+n_{Z})$ and
$g_{X}=\frac{n_{X}}{n_{X}+n_{Z}}$. Here we write
$n_{X}=N_{1,X_1}^{X}$, $n_{Z}=\sum_{l}N_{1,Z_{l}}^{Z}$ and
$e_1=e_{1,X_{1}}^{X,U}$ for simplicity. Note that $e_1^{p,Z,U}$ is a
function of $\langle s_0^X\rangle$. Straightly, we can also
formulate the upper bound of the phase-flip error rate of
single-photon counts in $X$ basis, being denoted by $e_1^{p,X,U}$.
We omit the explicit formula here since it is just trivially written
analogically to Eq.(\ref{eq:e1pZ}).

%where $\tilde{\tau}_{j}$ and $\hat{\tau}_{j}$ are defined in Eq.~(\ref{eq:TauLU}). In Eq.(\ref{eq:e1sta}), we have used $\langle t_0\rangle =\langle s_0\rangle/2$ deduced from the fact that the error rate of the dark count is $1/2$ as the dark counts occur randomly. Here $\langle t_0\rangle$ is the mean value of the error yield of the vacuum state.

Note that $\langle s_0^X\rangle$ (or $\langle s_0^Z\rangle$) is the
common variable in both quantity $\langle s_{1}^{X,L}\rangle $ and
$e_{1}^{p,Z,U}$  (or quantity $\langle s_{1}^{Z,L}\rangle $ and
$e_{1}^{p,X,U}$) shown in Eq.(\ref{ls1}) and
Eq.(\ref{eq:e1pZ}) respectively.  We need to know the range of this
for final key calculation.

If Alice uses a vacuum source in the protocol, the bounds are simply
\begin{equation}\label{eq:s0L0}
   \overline{S}_O^{\omega}=\langle s_{0}^{\omega,U}\rangle\geq \langle s_{0}^{\omega}\rangle \geq \langle s_{0}^{\omega,L}\rangle= \underline{S}_{O}^{\omega}.
\end{equation}

However, in practice, it is usually difficult to create a perfect
vacuum state in decoy-state QKD experiments~\cite{peng}. We may
consider the decoy-state methods without using a vacuum source. In
the  method, Eq.(\ref{eq:s0L0}) can not be used directly.

Reconsidering the relations in Eq.~(\ref{cons1}) with $j=1,2$, we can lower bound $\langle s_{0}^{\omega}\rangle$ by
eliminating $\langle s_{1,\alpha}^{\omega} \rangle$
\begin{equation}\label{eq:s0L1}
  \langle s_{0}^{\omega}\rangle \geq \langle s_{0}^{\omega,L}\rangle (\alpha)
  =\frac{a_{1,\alpha_{2}} \underline{S}_{\alpha_{1}}^{\omega} -a_{1,\alpha_{1}}
  \overline{S}_{\alpha_{2}}^{\omega}} {A_{\alpha_{1} \alpha_{2}}^{0,1}},
\end{equation}
where $A_{\alpha_{1} \alpha_{2}}^{0,1}=a_{0,\alpha_{1}}
a_{1,\alpha_{2}} -a_{0,\alpha_{2}} a_{1,\alpha_{1}}$. With these
preparations, in the method without the assumption of
vacuum, we can write the lower bound of $\langle
s_{0}^{\omega}\rangle$ as
\begin{equation}\label{eq:s0L}
  \langle s_{0}^{\omega}\rangle \geq \langle s_{0}^{\omega,L}\rangle =\max_{\alpha=X,Z}\left\{\langle s_{0}^{\omega,L}(\alpha)\rangle,0\right\}.
\end{equation}

By simply attributing all the errors to the vacuum pulses, we can upper bound $\langle s_{0}^{\omega}\rangle $ with
\begin{equation}\label{eq:s0U}
  \langle s_{0}^{\omega}\rangle \leq \langle s_{0}^{\omega,U}\rangle
  =\min \left\{ 2\overline {T}_{\omega_1}^{\omega}/a_{0,\omega_1},\overline{S}_{Z_1}^{\omega}/a_{0,Z_1},
  \overline{S}_{X_1}^{\omega}/a_{0,X_1}\right\}.
\end{equation}

\section{Numerical simulation} \label{sec:Simulation}
\subsection{Some special protocols}\label{subsec:PracticalDSM}
%As discussed in Sec.~\ref{sec:Protocol}, in order to obtain an nontrivial lower bound of $s_{1}^{\omega}$, we need at least two non-vacuum sources in the $X$ or $Z$ basis. On the other hand, in order to estimate the upper bound of $e_1^{X}$, we only need one non-vacuum source in the $X$ basis. Furthermore, we also need at least one source in the $Z$ basis to extract the final key from it. Here in this subsection, we list some special protocols with different number of sources.

Choosing different number of sources in the $X$ and $Z$ bases, we list some protocols in Tab.~\ref{tab:DiffMethods}.
\begin{table}[htbp]
%\begin{ruledtabular}
\begin{tabular}{c|c}
 \hline\hline
  {\textrm{Protocol}} & {\textrm{Description}}  \\
  \hline
  \textrm{3Int-0} & $O,Z_{1},Z_{2},X_{1},X_{2}$; ${Z_j}={X_j}$, $p_{Z_j}=p_{X_j}$, $q^Z=q^X$. \\
  \textrm{3Int-1} & $O,Z_{1},Z_{2},X_{1}$; $Z_1=X_1$, $Z_1\prec Z_2$, $\mu_{Z_2}=0.479$. \\
  \hline
  \textrm{4Int-1} & $O,Z_{1},Z_{2},X_{1}$; $Z_1\prec Z_2$. \\
  \textrm{4Int-2} & $Z_{1},Z_{2},X_{1},X_{2}$; $Z_1\prec Z_2$, $X_1\prec X_2$. \\
  \hline
  \textrm{5Int-1} & $O,Z_{1},Z_{2},X_{1},X_{2}$; $Z_1\prec Z_2$, $X_1\prec X_2$.\\
 \hline\hline
\end{tabular}
\caption{\label{tab:DiffMethods}List of some practical decoy-state methods. $O$ is the vacuum source. $Z_l$ and $X_r$ are the sources used by Alice in the $Z$ and $X$ bases respectively. The sources are different from each other except when we write the equivalent relations in the table, such as ${Z_j}={X_j}$. The probabilities of choosing the sources are independent except for the normalized condition. In 3Int-0, we also assume $Z_1\prec Z_2$. }
%\end{ruledtabular}
\end{table}

The first three-intensity protocol 3Int-0 listed in the table is the original symmetric method discussed in Ref.~\cite{peng}. In 3Int-0, Alice uses the vacuum source and two different non-vacuum sources in the $Z$ and $X$ bases to prepare the pulses. The symmetric conditions can be described by ${Z_j}={X_j}$, $p_{Z_j}=p_{X_j}$ for $j=1,2$ and $q^Z=q^X$ with the notations presented in this paper. In order to estimate the lower bound of the yield of single-photon pulses, we also need to assume $Z_1\prec Z_2$ which indicates that the sources $Z_1$ and $Z_2$ fulfill the relations presented in Eq.(\ref{eq:cond}) with $\sigma_{1}=Z_1$ and $\sigma_{2}=Z_2$.

The second three-intensity protocol 3Int-1 listed in Tab.~\ref{tab:DiffMethods} is the method considered in Ref.~\cite{MaSRep2013}. This is then further studied in Ref.~\cite{gm} with one more free intensity while the security loopholes as discussed earlier in this paper is still there and the key rate is not really fully optimized. In 3Int-1, Alice uses the vacuum source, two different sources $Z_1,Z_2$ in the $Z$ basis and only one source $X_1$ in the $X$ basis. The intensities of the source in the $X$ basis is equal to the first one in the $Z$ basis, i.e. ${Z_1}={X_1}$. Furthermore, the intensity (weak coherent state used in the simulation) of the second source in the $Z$ basis is $0.479$.

In Tab.~\ref{tab:DiffMethods}, we also list two different four-intensity protocols. The second one, 4Int-2,
is a new general protocol without the assumption of vacuum, whereas the vacuum source are used in the protocol 4Int-1.
Taking the protocol 4Int-2 as an example, Alice uses two different sources in the $Z$ and $X$ bases respectively.
For these four different sources, we assume ${Z_1}\prec {Z_2}$ and ${X_1}\prec {X_2}$.
In this protocol, we use sources $Z_2$ and $X_2$ to extract the final key. With Eq.(\ref{eq:s1L}),
the lower bounds of $s_{1,Z_{l}}^{Z}$ and $s_{1,X_r}^{X}$ can be estimated. The upper bound of $e_{1}^{p,Z}$ can be calculated with Eq.(\ref{eq:e1pZ}). The upper bound of $e_{1}^{p,X}$ can also be estimated in the same way. In 4Int-1, the sources $Z_2$ and $Z_1$ are used to extract the final key.

Besides these three-intensity and four-intensity protocols, we list a five-intensity protocol 5Int-1 in Tab.~\ref{tab:DiffMethods}.
In 5Int-1, Alice uses the vacuum source, two different sources in the $Z$ and $X$ bases to prepare the pulses. In order to make 5Int-1 contains 4Int-1 and 4Int-2 as its special cases, we need to introduce a probability $p_{Z_1}^{s}$ with
$0\leq p_{Z_1}^{s}\leq p_{Z_1}$. With probability $p_{Z_1}^{s}$, Alice will randomly choose pules from $Z_1$ to extract the final key.

To maximize the key rates, in 3Int-1 and 4Int-1, we shall use the following economic
worst-case estimation formula
\begin{equation}
  R = \min_{\langle s_0^X\rangle} \left[{R\left(\langle
  s_0^X\rangle\right)}\right]
\end{equation}
over the region for all possible values of $\langle s_0^X\rangle$ in
$[\langle s_0^{X,L}\rangle,\langle s_0^{X,U}\rangle]$.  Here
function $R$ is $R=R_1+R_2$ and \begin{equation}R_l=p_{Z_l}
q^Z\left\{a_{1,Z_l} s_{1,Z_l}^{Z,L} [1-H(e_1^{p,Z,U})]-f_e S_{Z_l}^{Z}
H(E_{Z_l}^{Z})\right\}.\end{equation} While in the
3Int-0, 4Int-2 and 5Int-1, we use the following economic more
worst-case estimation for key rates
\begin{equation}
  R = \min_{\langle s_0^Z\rangle,\langle s_0^X\rangle} \left[{R\left(\langle s_0^Z\rangle,\langle
  s_0^X\rangle\right)}\right]
\end{equation}
over the region for all possible values of $\langle s_0^Z\rangle$
and $\langle s_0^X\rangle$ in $[\langle
s_0^{Z,L}\rangle,\langle s_0^{Z,U}\rangle]$ and $[\langle
s_0^{X,L}\rangle,\langle s_0^{X,U}\rangle]$ respectively. Here
\begin{equation}
R(\langle s^X_0\rangle,\langle s^Z_0\rangle)=R_X+R_Z
\end{equation}
and
\begin{eqnarray}
  R_Z &=& p_{Z_2} q^Z\left\{a_{1,Z_2} s_{1,Z_2}^{Z,L} [1-H(e_1^{p,Z,U})]-f_e S_{Z_2}^{Z} H(E_{Z_2}^{Z})\right\}, \label{eq:RZ2} \\
  R_X &=& p_{X_2} q^X\left\{a_{1,X_2} s_{1,X_2}^{X,L} [1-H(e_1^{p,Z,U})]-f_e S_{X_2}^{X} H(E_{X_2}^{X})\right\}. \label{eq:RX2}
\end{eqnarray}
Note that in such a case, we need to calculate the final key rate with two variables, $\langle s_0^Z\rangle$
and $\langle s_0^X\rangle$ {\em jointly}. If not using this trick,
the simple worst-case treatment that treats each one separately will
produce a lower key rate.

 In subsection~\ref{subsec:Numerical},
we will compare the results for these protocols with the full
parameters optimization. After that, we will see that the final key
rates obtained with 4Int-2 and 5Int-1 are nearly equal to each
other. That is to say, it is advantageous to use the general
protocol 4Int-2 in practice, as it can give an almost optimal key
rate and has no use for the vacuum source.

\subsection{Numerical results}\label{subsec:Numerical}
In this subsection, we will present some results of the numerical simulation. We also optimize all parameters by the method of full optimization. For a fair comparison, we use all the sifted key corresponding to the successful events obtained with $X$ and $Z$ bases to extract the final key.
We shall estimate what values would be observed for the yields and error yields in the normal cases by the linear channel loss model shown in appendix~\ref{subsec:Model}. %For a fair comparison,
We use the same experimental parameters used in Ref.\cite{GobbyAPL2004} for our numerical simulation, which are also used for simulation in Ref.\cite{MaSRep2013}. The values of these parameters are listed in Table~\ref{tab:parameters}. In the simulation, we also assume that Alice uses the coherent states to prepare the pulses. Then the yields $S_{\alpha_j}^{\omega}$ and error yields $T_{\alpha_j}^{\omega}$ can be calculated with different intensities. By using these values, we can estimate the lower bound of $s_{1,Z_l}^{Z}(s_{1,X_r}^{X})$ and the upper bound of $e_{1}^{p,Z}(e_{1}^{p,X})$ with the method presented in Sec.~\ref{sec:Protocol}.

\begin{table}[htbp]
%\begin{ruledtabular}
\begin{tabular}{cccccccc}
 \hline\hline
  $e_d$ & $p_d$ & $\eta_d$ & $f_e$ & $\epsilon$ & $\alpha$  \\
  \hline
  $3.3\%$ & $1.7\times 10^{-6}$ & $4.5\%$ & $1.16$ & $10^{-10}$ & $0.2$ \\
 \hline\hline
\end{tabular}
\caption{\label{tab:parameters}List of experimental parameters used in numerical simulations. $e_d$: the misalignment-error probability, $p_d$: the background counting rate, $\eta_d$: the detector efficiency, $f_e$: the error correction inefficiency, $\epsilon$: the security bound considered in the finite-data analysis, i.e., failure probability, $\alpha$: the loss coefficient of the standard fiber measured in dB/km.}
%\end{ruledtabular}
\end{table}

To make a fair comparison, we make the full parameter optimization for all protocols. Here we also use the well-known local search algorithm~\cite{algorithm}. In this algorithm, we need to optimize the one-variable nonlinear function in each step for the local search.
In the optimization, except for those bits for error test, all bits are used for final key distillation. In particular, for protocols 3Int-1, 4Int-1, all bits due to sources $Z_1,Z_2$ are used for final key distillation; while in protocols 3Int-0, 4Int-2, bits due to sources $Z_2,X_2$ are used for final key distillation. In order to make the protocol 5Int-1 contains 4Int-1 and 4Int-2 as its special cases, besides bits due to sources $Z_2,X_2$ being used for final key distillation, we also need to split the bits due to source $Z_1$ into two parts (one for error test and the other for final key distillation) by introducing a probability $p_{Z_1}^{s}$ with $0\leq p_{Z_1}^{s}\leq p_{Z_1}$.

We consider the different methods in the case of data-size $N_t=10^{10}$. The optimal key rates of per pulse for the distances $80$km, $90$km, $100$km $110$km and $120$km (standard fiber), with the statistical fluctuations, are shown in Table~\ref{tab:Diffkms}.
Comparing the results with the original symmetric protocol 3Int-0 and the biased-basis protocol 3Int-1 discussed in Ref.~\cite{MaSRep2013}, the achievable key rate can be significantly improved with our new protocols.

\begin{table}[htbp]
%\begin{ruledtabular}
\begin{tabular}{c|ccccc}
 \hline\hline
  \hspace{12mm} & 80km & 90km & 100km & 110km & 120km  \\
  \hline
  3Int-0 & $3.37e$-$5$ & $1.73e$-$5$ & $7.65e$-$6$ & $2.39e$-$6$ & $3.91e$-$8$ \\
  3Int-1 & $5.04e$-$5$ & $2.38e$-$5$ & $9.29e$-$6$ & $2.02e$-$6$ & $0$ \\
  \hline
  4Int-1 & $5.43e$-$5$ & $2.65e$-$5$ & $1.09e$-$5$ & $2.99e$-$6$ & $3.65e$-$8$ \\
  4Int-2 & $5.05e$-$5$ & $2.39e$-$5$ & $9.63e$-$6$ & $3.50e$-$6$ & $4.55e$-$7$ \\
  \hline
  5Int-1 & $5.43e$-$5$ & $2.65e$-$5$ & $1.09e$-$5$ & $3.50e$-$6$ & $4.55e$-$7$ \\
 \hline\hline
\end{tabular}
\caption{\label{tab:Diffkms} Comparison of the optimal final key rates at different distances (standard fiber) with statistical fluctuation analysis in the case of data-size  $N_t= 10^{10}$(total number of pulses). Results for methods in Tab.~\ref{tab:DiffMethods} are listed here. Comparing with the results obtained with the original symmetric protocol 3Int-0 and the biased-basis protocol 3Int-1 presented in Ref.~\cite{MaSRep2013}, our protocols can significantly improve the the key rates. The key rate of our 4Int-2 protocol which does not use vacuum is very close to that of the 5Int protocol. Interestingly, the 5Int protocol automatically comes to one of the 4-intensity protocol at each distances. This strongly indicates that we have indeed reached the key rate maximization. This also shows that, for a given distance, instead of using the 5 intensity protocol, just choosing one of the 4 intensities is enough.}
%\end{ruledtabular}
\end{table}

In Tab.~\ref{tab:ParameterDiffkms}, we list the value of parameters with all of them being optimized. In the 1st column, we show the parameters for the original symmetric protocol 3Int-0. In the 2nd column, the parameters for the 4-intensity protocol 4Int-1 are listed. In the last column, we exhibit the parameters for our new 4-intensity protocol 4Int-2.
\begin{table}[htbp]
%\begin{ruledtabular}
\begin{tabular}{c|c|c}
  \hline\hline
  3Int-0 &  4Int-1 &  4Int-2 \\
  \hline
  \hspace{1mm} $p_{Z_2}=p_{X_2}=0.338$ \hspace{1mm} & \hspace{2mm} $p_{Z_2}=0.597$ \hspace{2mm} & \hspace{0mm} $p_{Z_2}=0.260$ \hspace{2mm}\\
  $p_{Z_1}=p_{X_1}=0.142$ & $p_{Z_1}=0.190$ & \hspace{2mm} $p_{Z_1}=0.077$ \hspace{2mm} \\
  $p_O=0.040$ & $p_{X_1}=0.112$ & \hspace{2mm} $p_{X_2}=0.458$ \hspace{2mm} \\
   & $p_{O}=0.101$ & \hspace{2mm} $p_{X_1}=0.205$ \hspace{2mm} \\
  \hline
  $\mu_{Z_2}=\mu_{X_2}=0.390$ & $\mu_{Z_2}=0.379$ & \hspace{2mm} $\mu_{Z_2}=0.419$ \hspace{2mm}  \\
  $\mu_{Z_1}=\mu_{X_1}=0.116$ & $\mu_{Z_1}=0.078$ & \hspace{2mm} $\mu_{Z_1}=0.200$ \hspace{2mm}  \\
  $\mu_{O}=0$ & $\mu_{X_1}=0.255$ & \hspace{2mm} $\mu_{X_2}=0.396$ \hspace{2mm} \\
   & $\mu_{O}=0$        & \hspace{2mm} $\mu_{X_1}=0.073$ \hspace{2mm} \\
  \hline
  $q^{X}=0.5$ & $q^{X}=0.223$ & \hspace{2mm} $q^{X}=0.579$ \hspace{2mm} \\
  \hline\hline
\end{tabular}
\caption{\label{tab:ParameterDiffkms} Comparison of parameters for the distance 110km (standard fiber) with statistical fluctuation analysis in the case of data-size $N_t= 10^{10}$ for different protocols. In the 1st, 2nd and 3rd columns we list the value of parameters with all of them being optimized for 3Int-0, 4Int-1 and 4Int-2 respectively. $p_{\alpha_j}$ is the probability to use the source $\alpha_j$ in the protocol. $p_{O}$ is the probability to choose the vacuum source. $\mu_{\alpha_j}$ is the intensity of the coherent source $\alpha_{j}$. $q^X$ is the probability that Bob measures the received pulses in the $X$ basis.}
%\end{ruledtabular}
\end{table}

\begin{figure}[htbp]
  \includegraphics[width=240pt]{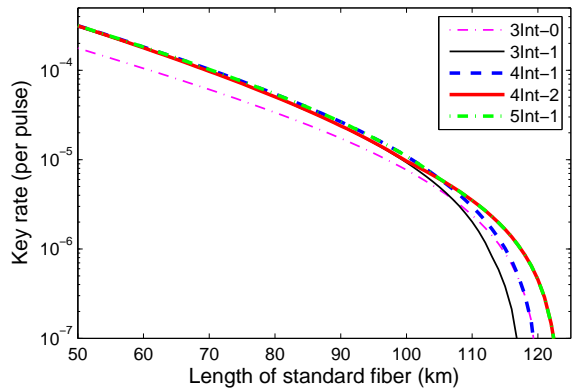}\\
  \caption{(Color online) Optimal secure key rates (per pulse) in logarithmic scale as as a function of the distance under different protocols. Here we set $N_t=10^{10}$. The key rates are obtained by using numerical methods with 3Int-0 (thin magenta dash-dot line), 3Int-1 (thin black solid line), 4Int-1 (blue dashed line), 4Int-2 (red solid line) and 5Int-1 (green dash-dot line). The key rates with four-intensity protocol 4Int-2 and five-intensity protocol 5Int-1 are almost overlapped. In simulation, we perform a full parameter optimization for all cases.}\label{fig:CompareNt9}
\end{figure}

More extensive results are shown in Fig.~\ref{fig:CompareNt9}. In this figure, we show the optimal key rate (per pulse) in logarithmic scale as
a function of the distance under a practical setting with finite data-set $N_t=10^{10}$.
In Fig.~\ref{fig:CompareNt9}, we use the thin magenta dash-dot line, the thin black solid line, the blue dashed line, the red solid line and the green
dash-dot line to indicate the results obtained with the protocols 3Int-0, 3Int-1, 4Int-1, 4Int-2 and 5Int-1 respectively. From the numerical simulation,
we can conclude that both the achievable key rate and distance can be significantly improved by using our new protocols compared with 3Int-0 and 3Int-1.
The optimal final key rates obtained by using the four-intensity protocol 4Int-2 and the five-intensity protocol 5Int-1 are nearly equal to
each other. In practical QKD applications, for better performance in terms of key rate and distance, it is advantageous to make use of 4Int-2 with
full parameter optimization, because of the advantage of not using any vacuum source.
%It should be note that the optimal key rate obtained with 4Int-1 where the vacuum source is used is a little higher than that obtained with 4Int-2 when we consider the situation in which the distance approaches to the limits (127km in Fig.~\ref{fig:CompareNt9}). The reason is that the successful events related to the vacuum state can be treated together in the statistical analysis as the vacuum state is independent of the basis used by Alice to prepare the pulses.

\begin{figure}[htbp]
  \includegraphics[width=240pt]{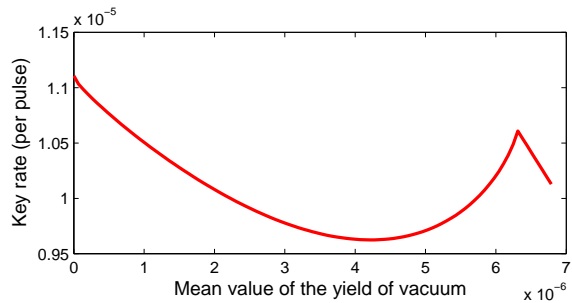}\\
  \caption{(Color online) The key rate $R(\langle s_0^{X}\rangle)=R(\langle s_0^{X}\rangle,\langle \tilde{s}_0^{Z}\rangle)$ vs the yield of vacuum state as the variable and $\langle s_0^{X}\rangle \in [\langle s_0^{X,L}\rangle, \langle s_0^{X,U}\rangle]$ with 4Int-2 for the distance 100km (standard fiber). From this figure, we can see that the smallest values of $R$ is not obtained at the lower bounds of $\langle s_0^{X}\rangle$. }\label{fig:RofS0}
\end{figure}

As discussed in Sec.~\ref{sec:Protocol}, in order to estimate the final secure key rate, we need to find out the smallest one with variable $\langle s_0^{X}\rangle$ changing in the region $[\langle s_0^{X,L}\rangle, \langle s_0^{X,U}\rangle]$. Previously, in Refs.~\cite{LoJC2005,MaSRep2013}, they straightly use the lower bound $\langle s_0^{X,L}\rangle$ to calculate the final key rate. In order to insure the security, we must make sure that the smallest key rate value just happened at the lower bound value of $\langle s_0^{X}\rangle$ when we use it to calculate the final key rate directly. However, in the numerical simulation, we find some counterexamples. In Fig.~\ref{fig:RofS0}, after evaluating $\langle s_0^{Z}\rangle$ with its proper value $\langle \tilde{s}_0^{Z}\rangle$, we plot the key rate $R$ as a function of $\langle s_0^{X}\rangle$ with $\langle s_0^{X}\rangle \in [\langle s_0^{X,L}\rangle, \langle s_0^{X,U}\rangle]$ for the protocol 4Int-2 at the distance of 100km. From this figure, we can see that the smallest value of $R(\langle s_0^{X}\rangle)=R(\langle s_0^{X}\rangle,\langle \tilde{s}_0^{Z}\rangle)$ does not obtained with $\langle s_0^X\rangle=\langle s_0^{X,L}\rangle$. In this case, we can not calculate the final key rate with $\langle s_0^{X,L}\rangle$ directly. That is to say, if we use the lower bound of the yield of vacuum state to calculate the final key rate in this counterexample, the security will not be guaranteed.

\section{Conclusion}\label{sec:Conclusion}
We propose to make use of the yields of those pulse which are prepared in one basis and measured in another basis. This is to say, we can treat those measurement outcome from different preparation bases {\em jointly} provided that they are measured in the same basis. Based on this, we present correct formulas for the decoy-state method with considering the effects of statistical fluctuations.

In practice, it is usually difficult to create a perfect vacuum state in decoy-state QKD experiments. In this paper, we also present an analytical approach with general decoy states, i.e. without the assumption of vacuum. By taking the full parameter optimization with the linear channel loss model, we find that our new 4-intensity protocol 4Int-2 can give an almost optimal key rate. The vacuum source and more decoy states cannot help to increase the key rate.

In the simulation, we find some counterexamples in which the lower bound of the final key rate does not obtained with the lower bound of the yield of the vacuum state. That is to say, if we use the lower bound of the yield of the vacuum state to calculate the final key rate directly in the counterexample, the security will not be guaranteed. So we need to find out the smallest key rate with different values of the yield of the vacuum state faithfully.

{\bf{Acknowledgement:}} We acknowledge the financial support in part by the 10000-Plan of Shandong province (Taishan Scholars), the National High-Tech Program of China, Grants No. 2011AA010800 and No. 2011AA010803, and the NSFC Grants No. 11474182, No. 11174177, and No. 60725416, the Open Research Fund Program of the State Key Laboratory of Low-Dimensional Quantum Physics Grant No. KF201513, and the Key R\&D Plan Project of ShanDong Province, Grants No. 2015GGX101035.

\appendix
\section{Linear Channel Loss Model}\label{subsec:Model}
In order to do the numerical simulations, we need the observed yields and error yields. Here we consider a widely used linear channel loss model~\cite{wang05}. With this model, the yields and error yields can be simulated. In this model, Alice randomly chooses the states for different sources with corresponding probabilities.

For this model, we define $\eta^{\omega}$ as the overall transmission of the channel when Bob measures the pulses in the $\omega$ basis ($\omega=X,Z$). It should be note that $\eta^{\omega}$ includes the detector efficiency and is measured in dB. In the real setups, such as in the situation with different detector efficiencies, the transmissions $\eta^{X}$ and $\eta^Z$ are not equal to each other rigorously. So we should treat the yields separately with different bases used by Bob to measure the pulses.

It is reasonable to assume independent between the behaviors of the $k$ photons in a $k$-photon state. Therefore the transmittance of the $k$-photon state $\eta_{k}^{\omega}$ is given by
\begin{equation}\label{eq:etak}
  \eta_{k}^{\omega}=1-(1-\eta^\omega)^k, \quad (k=0,1,2,\cdots)
\end{equation}
when Bob measures the pulses in the $\omega$ basis.

Define $s_{k,\alpha}^{\omega}$ to be the yield of an $k$-photon state when Alice prepares the state in the $\alpha$ basis and Bob measures in the $\omega$ basis ($\omega,\alpha=X,Z$). Specially, $s_{0,\alpha}^{\omega}$ is the background rate. It should be note that $s_{0,\alpha}^{\omega}$ is independent of the $\alpha$ basis used by Alice to prepare the pulses. Then we can write $s_{0,\alpha}^{\omega}$ into $s_{0}^{\omega}$ concisely.

As discussed above, we need to treat $s_{k,\alpha}^{\omega}$ with $\omega=X$ and $\omega=Z$ separately. In the protocol, there are two detectors are used when Bob chooses the $\omega$ basis to measure the pulses. Furthermore, Bob denote the situation when one and only detector makes a count as a successful event. The yield of an $k$-photon state, $s_{k,\alpha}^{\omega}$, mainly comes from two parts, background and true signal. Assuming that the background counts are independent of the signal photon detection, then $s_{k,\alpha}^{\omega}$ is given by
\begin{equation}\label{eq:skSame}
  s_{k,\alpha}^{\omega}=(1-s_{0}^{\omega})[1-(1-2s_{0}^{\omega})(1-\eta^\omega)^k],
\end{equation}
when Alice and Bob choose the same bases, i.e. $\omega=\alpha$. Furthermore, we have
\begin{equation}\label{eq:skDiff}
  s_{k,\alpha}^{\omega}=2(1-s_{0}^{\omega})[(1-\eta^{\omega}/2)^k -(1-s_{0}^{\omega})(1-\eta^{\omega})^k],
\end{equation}
when the bases used by Alice and Bob are different, i.e. $\omega\neq \alpha$.

In the ideal case without misalignment error, the error rate of the $k$-photon state when Alice and Bob use the same bases, $\hat{e}_{k,\omega}^{\omega}$, is given by
\begin{equation}\label{eq:ekSame}
  \hat{e}_{k,\omega}^{\omega}=s_{0}^{\omega}(1-s_{0}^{\omega})(1-\eta^{\omega})^k.
\end{equation}
In the real setups, we denote $e_d$ as the misalignment error probability. Then the error rate of the $k$-photon state is
\begin{equation}\label{eq:ekSameed}
  e_{k,\omega}^{\omega}=e_d(1-2\hat{e}_{k,\omega}^{\omega})+\hat{e}_{k,\omega}^{\omega}.
\end{equation}

With these notations, we can write the yields of the pulses prepared from source ${\alpha_j}$ and measured in the ${\omega}$ basis as
\begin{equation}\label{eq:SSame}
  S_{\alpha_j}^{\omega}=\sum_{k}{a_{k,\alpha_j} s_{k,\alpha}^{\omega}},
\end{equation}
where the density matrix of source $\alpha_j$ given by Eq.(\ref{eq:rho}) has been used. The error yield is given by
\begin{equation}\label{eq:ESame}
  T_{\alpha_j}^{\omega}=E_{\alpha_j}^{\omega}S_{\alpha_j}^{\omega}=[e_d(1-2\hat{E}_{\alpha_j}^{\omega})+\hat{E}_{\alpha_j}^{\omega}] S_{\alpha_j}^{\omega},
\end{equation}
when $\omega=\alpha$ with $\hat{E}_{\omega_j}^{\omega}=\sum_{k}{a_{k,\omega_j} e_{k,\omega}^{\omega}}$. Furthermore, the error rate of the $k$-photon state is always equal to $1/2$ when Alice and Bob use the different bases.

If we consider the weak coherent sources, assuming that the phase of each pulse is totally randomized, the density matrix of the coherent state with intensity $\mu$ can be written as
\begin{equation*}
  \rho=\sum_{k}\frac{e^{-\mu}\mu^k}{k!} \oprod{k}{k}.
\end{equation*}
With this formula, the yield of the pulses prepared from the weak coherent source ${\alpha_j}$ with density $\mu_{\alpha_j}$ and measured in the $\omega$ basis is
\begin{equation}\label{eq:SWCSame}
  S_{\alpha_j}^{\omega}=(1-s_{0}^{\omega})[1-(1-2s_{0}^{\omega})e^{-\mu_{\alpha_j} \eta^{\omega}}],
\end{equation}
when $\omega=\alpha$ and
\begin{equation}\label{eq:SWCDiff}
  S_{\alpha_j}^{\omega}=2(1-s_{0}^{\omega})e^{-\frac{\mu_{\alpha_j} \eta^{\omega}}{2}}[1-(1-s_{0}^{\omega})e^{-\frac{\mu_{\alpha_j} \eta^{\omega}}{2}}],
\end{equation}
when $\omega\neq \alpha$. The error yield can be simulated by Eq.(\ref{eq:ESame}) when $\alpha=\omega$ with $\hat{E}_{\omega_j}^{\omega}=s_{0}^{\omega}(1-s_{0}^{\omega})e^{-\mu_{\omega_j}\eta^{\omega}}$.

%\clearpage

\end{document}